\def   \ni {\noindent}

\def   \ssk {\vskip  5truept}

\def   \bsk {\vskip 15truept}
 
\def   \newpage {\vfill\eject}
\def   \newline {\hfil\break}

\def\deg{$^{\circ}$}

\def\la{\mathrel{\mathchoice {\vcenter{\offinterlineskip\halign{\hfil
$\displaystyle##$\hfil\cr<\cr\sim\cr}}}
{\vcenter{\offinterlineskip\halign{\hfil$\textstyle##$\hfil\cr
<\cr\sim\cr}}}
{\vcenter{\offinterlineskip\halign{\hfil$\scriptstyle##$\hfil\cr
<\cr\sim\cr}}}
{\vcenter{\offinterlineskip\halign{\hfil$\scriptscriptstyle##$\hfil\cr
<\cr\sim\cr}}}}}

\documentstyle[epsfig]{article}
\begin{document}

\hsize 5truein
\vsize 8truein
\font\abstract=cmr8
\font\keywords=cmr8
\font\caption=cmr8
\font\references=cmr8
\font\text=cmr10
\font\affiliation=cmssi10
\font\author=cmss10
\font\mc=cmss8
\font\title=cmssbx10 scaled\magstep2
\font\alcit=cmti7 scaled\magstephalf
\font\alcin=cmr6 
\font\ita=cmti8
\font\mma=cmr8
\def\ref{\par\noindent\hangindent 15pt}
\null



\title{\ni COMPACT AND DIFFUSE SOURCES IN THE GALACTIC CENTER REGION}

\bsk \bsk
\author{\ni E.~Churazov $^{1,2}$, M.~Gilfanov $^{1,2}$, R.~Sunyaev
$^{1,2}$, S.~Kuznetsov $^{2}$}                                                        
\bsk
\affiliation{1) MPI fur Astrophysik,
Karl-Schwarzschild-Strasse 1, 85740 Garching, Germany} 
\ssk
\affiliation{2) Space Research Institute (IKI), Profsouznaya 84/32,
Moscow 117810,  Russia}                                                
\bsk
\baselineskip = 12pt

\abstract{ABSTRACT \ni
The 10 by 10 degrees region around dynamic center of our Galaxy is
known to host a large number of bright X--ray sources, most of which are low
mass binary systems. While high luminosity Z--sources are especially
bright in the standard X--ray band (e.g. 2--10 keV), lower luminosity
systems are dominating the flux above 30 keV. The two hardest sources
in the region, i.e. 1E1740.7--2942 and GRS 1758--258 are thought to be black
hole candidates (BHC) based on the similarity of spectral properties and
short--term variability to that of dynamically proven BHCs. However, while
most of the other galactic BHCs are known to be transients, these two
objects could be considered persistent though variable
sources. The persistent behavior of these sources implies some
constraints on the parameters of the binary systems. Both sources were
detected in the first GRANAT observation of the GC region. They are 
the only two persistent 
BHCs seen in the 35-75 keV SIGMA image after collecting 
several million seconds of exposure time over eight years,
indicating that they are perhaps the only two persistent BHCs in the region. 

Another peculiar object in the field is of course a putative
supermassive black hole (Sgr A*) at the dynamic center of our Galaxy.
The present day's X--ray luminosity from Sgr A* is very small ($\la
10^{36}$ erg/s). However
an indication that Sgr A* might have been much brighter in the past
was found while studying diffuse X--ray emission in the region. We
speculate here on how future X--ray observatories may verify the hypothesis
of the violent activity of Sgr A* in the past. 
}                                                    
\bsk
\baselineskip = 12pt

\bsk
\baselineskip = 12pt


\text{\ni 1. COMPACT SOURCES
\ssk
\ni

The Galactic Center (hereafter GC) is a densely populated region of the
sky. Few dozen bright sources are known within
5-6 degrees from the GC. A large fraction of these sources are low mass
binary systems (LMXB) containing a neutron star or a black hole. Shown in
Fig.1 are the images taken in the standard (2-20 keV, TTM/KVANT, see
also Ubertini et al., 1999 for the BeppoSAX/WFC results) and
hard (35-75 keV, SIGMA/GRANAT) energy bands. Both images have a similar dynamic
range (flux ratio of the brightest and weakest sources shown in the
image) of $\sim$10--20 and contain a comparable number of bright
sources. Remarkably, the subsamples of sources seen in the two
energy bands are very different. In fact the X--ray burster GX354--0 is
the only source seen as a bright object in both images. The reason for
that is clear from Fig.2, where the spectra of typical LMXB sources
\newpage

\noindent from the GC region are shown. High luminosity LMXBs with a neutron
star (Z--sources, Fig.2 left) are very bright in the standard X--ray
band, but are very weak above $\sim$30 keV. On the contrary BHCs
(Fig.2 right) have peak luminosity at the energy of $\sim$100
keV, while in standard X--ray band they are relatively
weak. An intermediate group is 
constituted by lower luminosity LMXB with a neutron 
star (typically X--ray bursters, Fig.2 middle), which are often switched
between softer and harder states. The latter sources appear part of
the time as bright objects in the standard X--ray band and part of
the time as bright objects in the hard band. Even in the hard state
spectra of X--ray bursters are still softer (in terms of the energy at
which maximum luminosity is emitted) than those of BHCs (e.g. Goldwurm et
al. 1994, Gilfanov et al. 1995, Churazov et al., 1997). Indeed,
using the optically thin plasma emission model to crudely characterize
hardness of source spectra in the 35-150 keV band, one can conclude
that best fit temperatures to the spectra of each NS LMXB seen in the 
SIGMA image (Fig.1) are clustered around 50 keV or below, while
best fit temperatures for BHCs (e.g. 1E1740.7-2942, GRS1758-258) are all around
100 keV or above.  

\begin{figure}[t]
\hfil\psfig{file=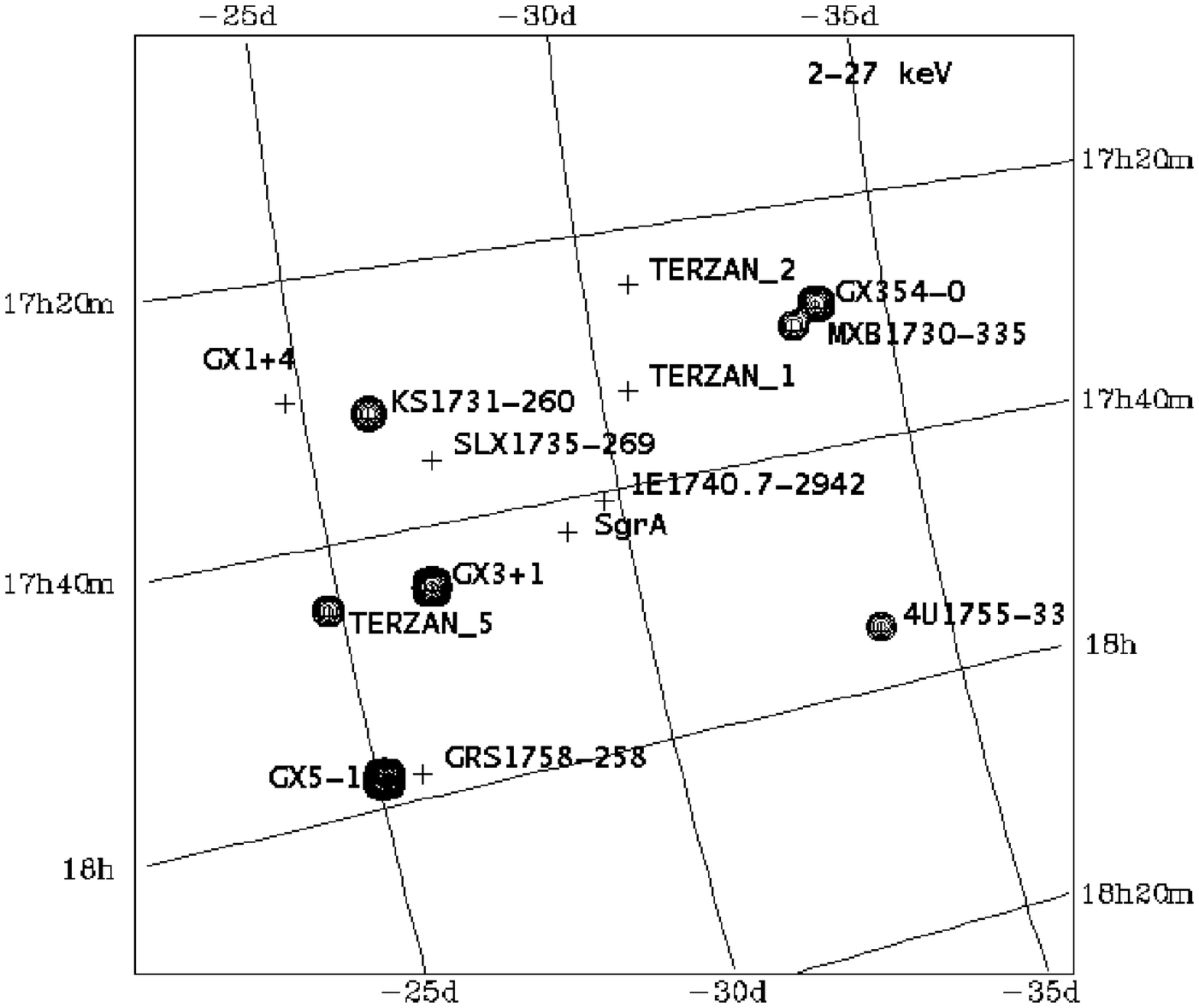, width=5.2cm}\hfil
\psfig{file=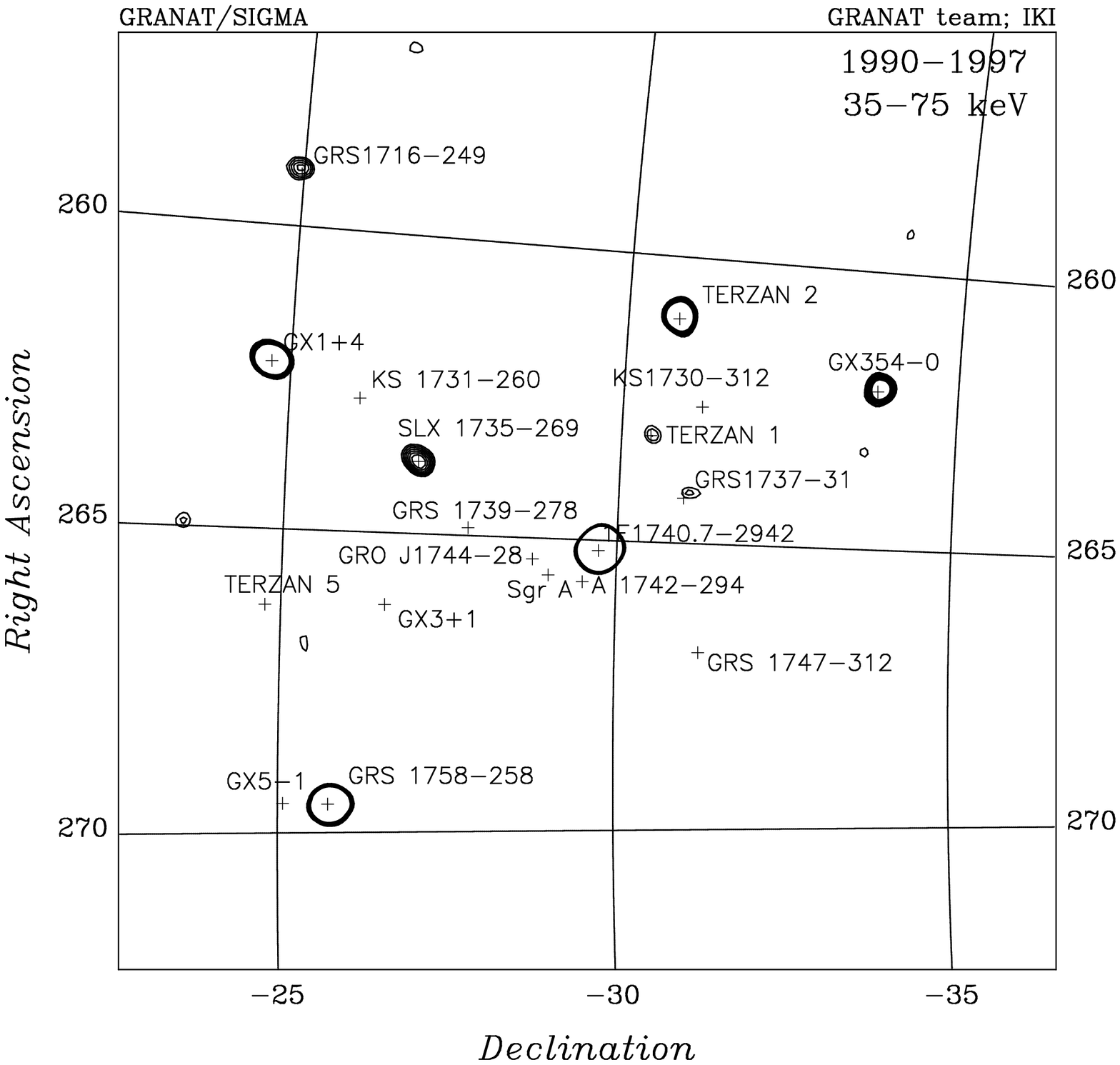,width=6.6cm,angle=270}

\caption{
FIGURE 1. The Galactic Center region in the standard 2-27 keV
(TTM/KVANT) and hard 35-75 keV (SIGMA/GRANAT) bands. Both images have
roughly the same dynamic range (flux ratio of the brightest and
weakest sources shown in the image) of $\sim$10--20. The exposure time is
$\sim$4000 s and $\sim$6 10$^6$ s for the TTM and SIGMA images
respectively. For the standard X-ray band the image was cut below
5$\sigma$ level (corresponding to the flux of $\sim$ 15 mCrabs at the
center of the image) and for the hard band the image was cut below
4.5$\sigma$ level (corresponding to the flux of $\sim$ 5 mCrabs at the
center of the image).  
} 
\label{gcimage}
\end{figure}

The 1E1740.7-2942 and GRS1758-258 are the hardest sources in the
field, having peaks of the spectra ($\nu F_\nu$ in units
$keV^2~cm^{-2}~s^{-1}~keV^{-1}$) at the energy of $\sim$100 keV. Their
spectral properties resemble those of the
dynamically proven BHCs. However while most of the BHCs in the Galaxy
are transients, 
1E1740.7-2942 and GRS1758-258 are certainly persistent though
variable sources. Shown in Fig.3 are the 40--150 keV light curves of
1E1740.7-2942 and the transient GRS1716-249 (Nova Oph 93 -- also located
in the GC region) for most of SIGMA observations in 
1990--1998. It is clear that the light curve of 1E1740.7-2942 is much 
closer to the light curve of Cyg X--1 (see e.g. Kuznetsov et al.,
1997), than that of a typical X--ray transients. The same is
true for GRS1758-258. However Cyg X-1 has a very massive companion
and accretion is fed by a stellar wind. At least for GRS1758-258 the
hypothesis of the companion massive enough to produce sufficient
stellar wind can be rejected based on the optical and infrared limits
(see e.g. Chen, Gehrels \& Leventhal, 1994; Marti et al., 1998). 

\begin{figure}
\hfil\psfig{file=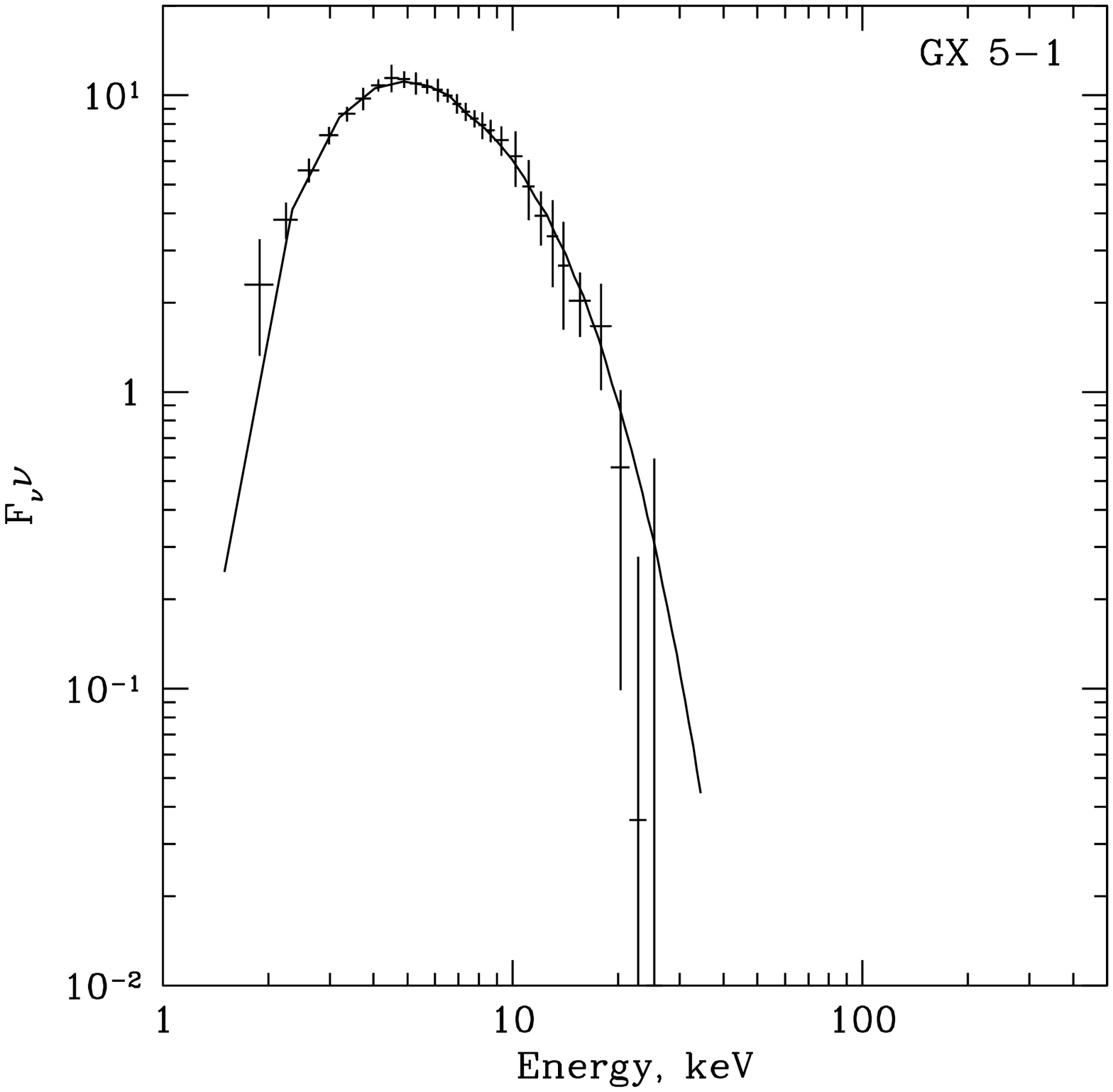,
width=3.8cm}\hfil\psfig{file=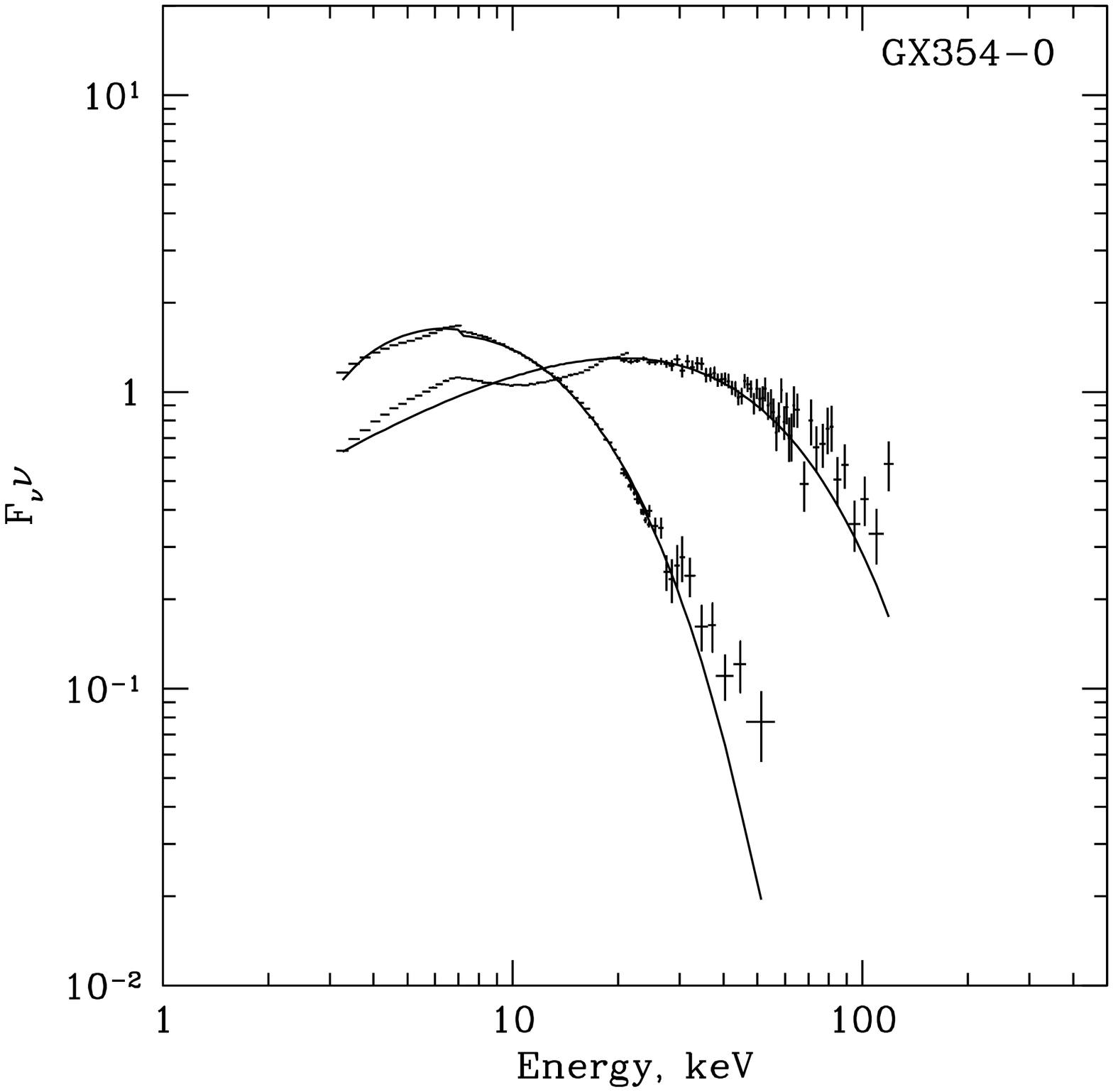,
width=3.8cm}\hfil\psfig{file=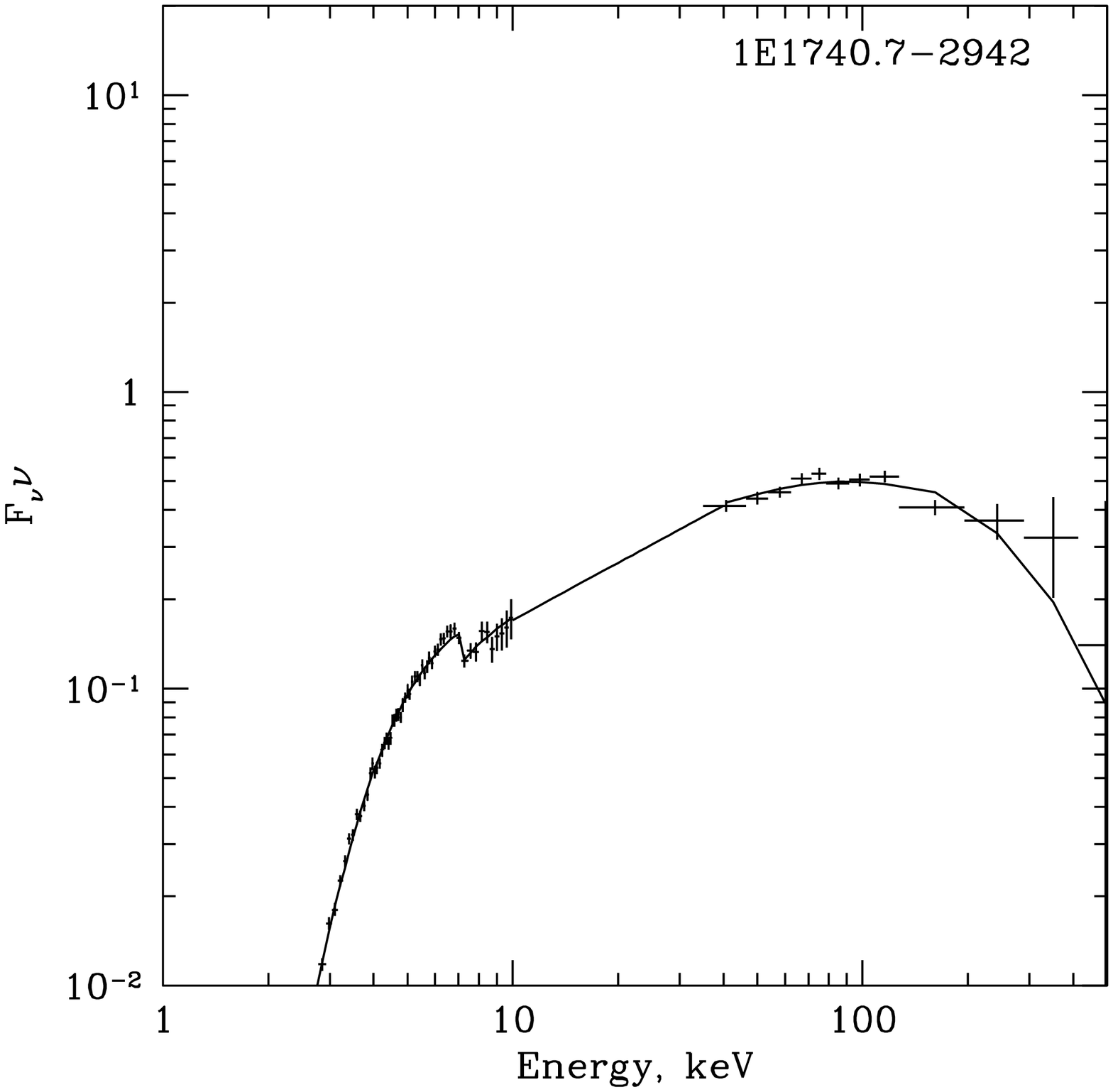, width=3.8cm}
   
\caption{FIGURE 2. Typical spectra ($\nu F_\nu$ in units
$keV^2~cm^{-2}~s{-1}~keV^{-1}$) of high luminosity neutron star LMXB GX 5--1
(left, TTM/KVANT), X--ray burster GX354-0 in two spectral states
(middle, RXTE) and 
black hole candidate 1E1740.7--2942 (right, ASCA, SIGMA/GRANAT). 
} 
\label{3spec}
\end{figure}

During the last few years the hypothesis explaining the ``transient''
character of the black hole LMXBs due to the thermal--viscous
instability of an accretion disc in the zone of partial hydrogen
ionization was intensively discussed in the literature (see e.g. Van
Paradijs 1996, King et al. 1997a, Dubus et al. 1998). This instability
was successfully applied to cataclysmic variables (e.g. Meyer 
and Meyer-Hofmeister 1981). It is interesting to understand the role
of this instability for the GRS~1758-258 and 1E1740.7-2942 sources
(Kuznetsov et al., 1998). The
light curves of GRS~1758-258 and 1E1740.7-2942 seem to indicate that
this instability is suppressed in these objects. Menou, Narayan \&
Lasota (1998) suggested that a population of faint non-transient
LMBHBs can exist in the Galaxy. In their non-transient sources the
outer regions of the disk are colder than the ionization temperature
of hydrogen while at smaller radii disk switches to the stable advection
dominated solution before the disk temperature rises to the critical
value. However expected luminosities of such objects are much smaller
than observed values ($\ge 10^{37}~erg/s$) for GRS~1758-258 and
1E1740.7-2942. Another possibility is to assume that even at the
outmost regions of the accretion disk the temperature is already above
the hydrogen ionization temperature. Unlike CVs for accreting black holes the
irradiation of the outer part of the accretion disc by the X-rays
(e.g. Shakura \& Sunyaev 1973, Lyuty \& Sunyaev 1976) originated in
the inner part of the disk 
might be important for suppressing this instability (e.g. Van
Paradijs 1996, King et al. 1997a, 1997b, Dubus et al. 1998). Using their
prescriptions for suppression of the instability (in the form of
minimal mass accretion rate as a function of radius) one can place a
constraint on the orbital period of the binary, assuming that outmost
radius of the disk corresponds to some fraction of the primary Roche
lobe radius. For reasonable choice of the black hole mass and mass ratio the
period must be less than 10-20 hours  and the
mass accretion rate in GRS1758-258 and 1E1740.7-2942 should be in the
range $\sim 10^{17}$ -- $10^{18}~g/s$. The above estimate is based on
the results of calculations of Dubus et al., 1998. Note however that
there are many
uncertain parameters in the model which might affect these
estimates. Use of the prescription of King et al. 1997b leads to a
more tight constraint on the binary period ($P_{orb}\le$ 7
hours). The assumption that only local viscous energy release affects
the temperature and the structure of the disc (i.e. no irradiation)
imply even stronger limit on the period $P_{orb}\le$ 3-5 hours. Short
period BH transients are known (e.g. GRO J0422+32 has $P_{orb}=$ 5.1
hours), but it is interesting that 
mass accretion rate of $\sim 10^{17}$ -- $10^{18}~g/s$ 
(sustained  by GRS1758-258 and 1E1740.7-2942 at approximately the same
level for at least 10--20 years) is two orders of magnitude
higher than the averaged mass accretion rate for transient sources (see
e.g. Van Paradijs 1996). This may be related with the different type of 
companion stars in GRS1758-258 and 1E1740.7-2942 and BHC transients. 
On the other hand many persistent neutron star binaries are know with
luminosities in the range of $10^{37}$ -- $10^{38}~erg/s$
(i.e. $\dot{M}\sim 5~10^{16}~-~5~10^{17}~g/s$). Using this empirical
fact one can further speculate that  GRS1758-258 and 1E1740.7-2942 might be
the systems with low masses of the black holes, compared to the higher
black hole masses in the transient systems.

\begin{figure}[t]
\centerline{\psfig{file=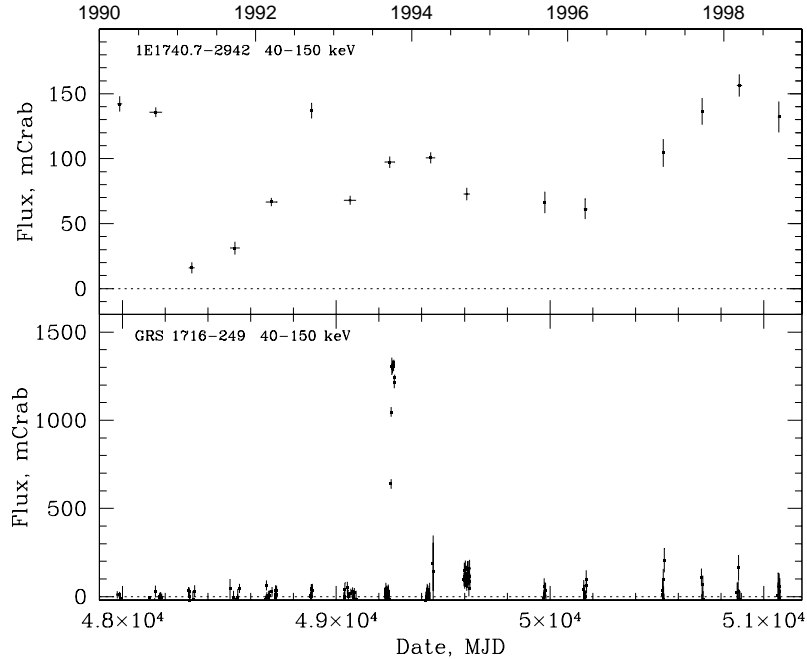, clip=, width=11cm}}
\caption{FIGURE 3. The long term light curves of two black hole candidates in
the Galactic Center region. GRS 1716--249 (X--ray Nova Oph 93) is
a transient source with a small duty cycle. On the contrary
1E1740.7--2942 was radiating in the hard X--rays at approximately the
same level during most of the SIGMA/GRANAT observations in
1990--1998. Another non--transient black hole candidate in the region
-- GRS1758-258 has a light curve similar to that of 1E1740.7--2942.
} 
\label{lc2}
\end{figure}

It is interesting that 1E1740.7-2942 and GRS1758-258 were
detected in the very 
first GC observation by SIGMA in 1990. But even after adding more than
100 observations over 1990--1998 these two objects remain the only
persistent BHCs in the GC field, although the sensitivity increased by
a factor of $\sim$10. The only other BHC seen in the
averaged 1990--1998 image is GRS1716--249 (an X--ray
transient Nova Oph 93), which was bright for only a short period of
time. This result suggests that persistent black holes are rare objects
and even with increased INTEGRAL sensitivity 1E1740.7-2942 and
GRS1758-258 may remain the only two persistent BHCs in the
field. The absence of weaker BHCs in the GC field may also imply some
constraints on the luminosity function of the persistent black hole
candidates in the hard state.

\bsk
\ni 2. DIFFUSE EMISSION -- EVIDENCE FOR SGR A* ACTIVITY 
\ssk
\ni 
While at least three stellar mass BHCs are seen in the averaged 35--75
keV SIGMA image (Fig.1) there is no evidence for strong X--ray
emission from the dynamic center of the Galaxy. The nonthermal radio source Sgr
A*, which presumably coincides with a supermassive ($\sim
2.5~10^6~M_\odot$, e.g. Eckart \& Genzel, 1996) black hole is a very
weak object in X--rays. The 
best X--ray imaging  of the Sgr A* vicinity so far has been done
with 
ROSAT (Predehl \& Truemper, 1994) resolving the 1E1742.5-2859 source (usually
associated with Sgr A* X--ray emission) into three components, one of
which coincides (within $10''$) with the 
position of the radio source Sgr A* (See also Sidoli et al., 1999 for
the results of recent deep observations of BeppoSAX). The intrinsic
luminosity of the source in this energy band (1.2--2.5 keV) is
uncertain because of the strong interstellar absorption and ranges
from less than $10^{35}$ to $7~10^{35}$ 
erg/s, depending on the adopted value of the absorbing column
density. At higher energies (2-200 keV) the absorption is not so 
important, but the spatial resolution
(typically few arcminutes) is significantly worse than that of ROSAT
and detected fluxes may be contaminated by other sources. In any case the
conservative upper limit on the X--ray luminosity of the Sgr A* does
not exceed few $10^{36}$ erg/s. This value is some $10^8$ times lower
than the Eddington limit for a $\sim 2.5~10^6~M_\odot$ black hole. 

While the present day X--ray luminosity of Sgr A* is small, some
indication of a past violent activity of the Sgr A* were
found. Observations of the diffuse emission in the 8--22 keV energy 
range, elongated parallel to the Galactic plane (Sunyaev et al., 1993) and 
detection of the strong 6.4 keV fluorescent line with $\sim$ 1 keV 
equivalent width from Sgr B2 complex and few other clouds in the
Galactic Center region (Koyama, 1994, 1996) suggest that the neutral matter of
these clouds is (or was) illuminated by powerful X-ray radiation,
which gave rise to the reprocessed radiation. It is important to
stress that the large value of the 6.4 keV line equivalent width detected
by ASCA from the Sgr B2 cloud suggests that we observe an almost pure
reprocessed component without direct emission from the source. It is
therefore likely that this source is now in a weak state. One of the
interesting hypothesis suggests that reprocessed radiation was
produced by a short outburst of the X-ray emission from Sgr A* few
hundred years ago. Simple estimates show that Sgr A* luminosity at the
level of $10^{39}$ erg/s is required to power fluorescent emission
from the Sgr B2 cloud (i.e. well in the range allowed by the Eddington
limit for this object). We consider below a few simple observational tests
which could verify this hypothesis.

\begin{figure}[t]
\hfil\psfig{file=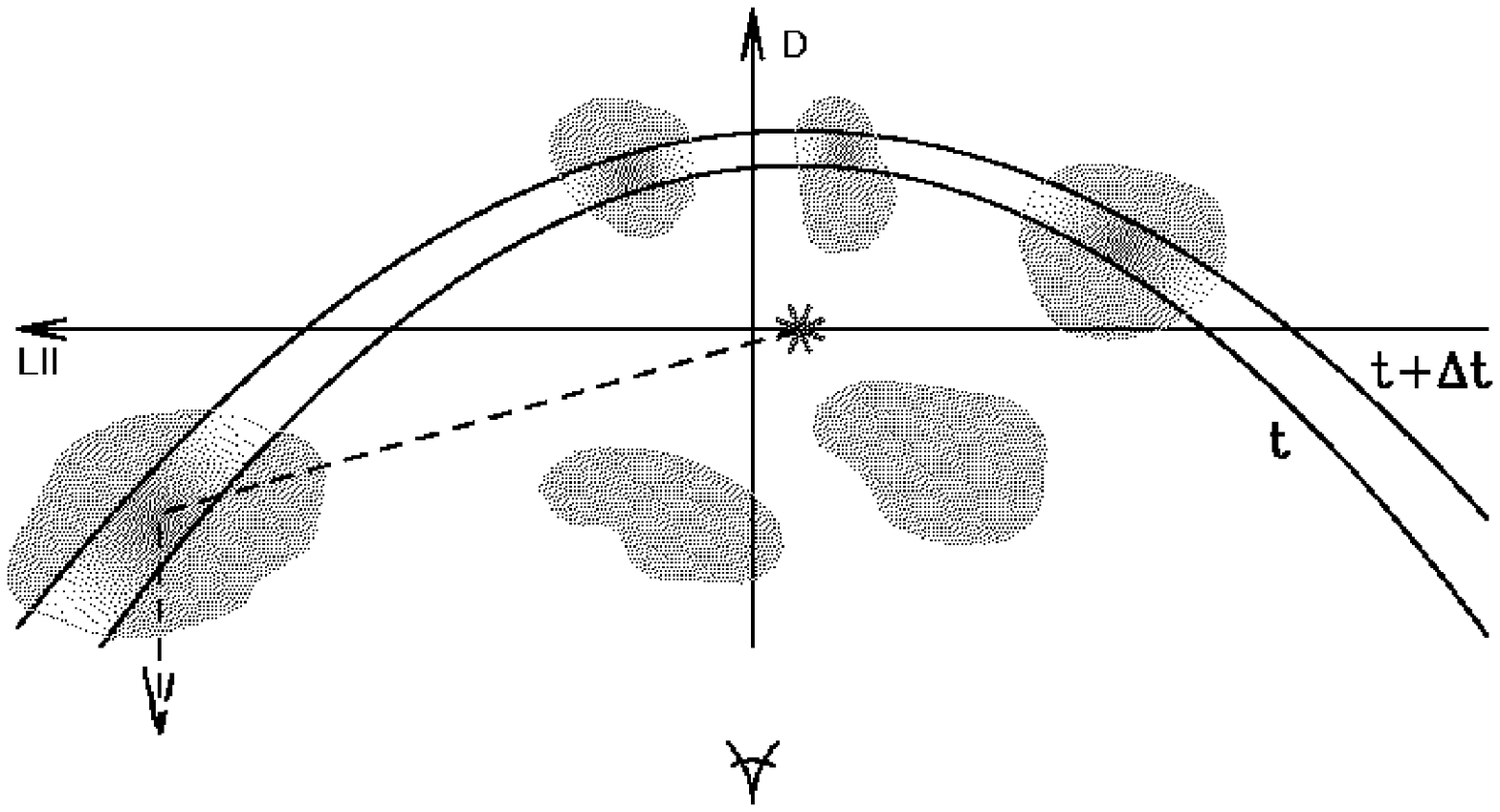, clip= ,width=5.6cm}
\hfil\psfig{file=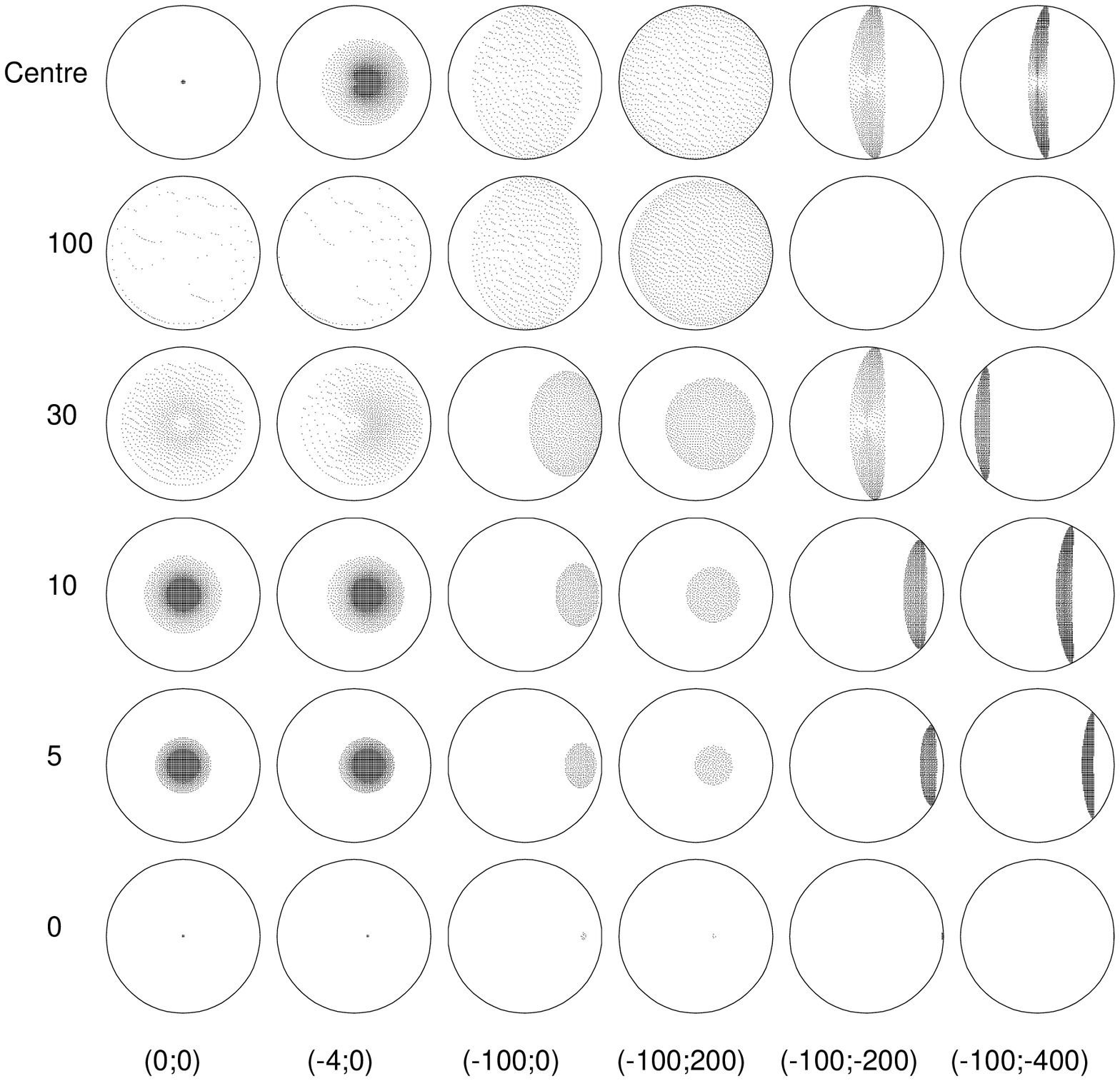, clip=, width=5.6cm}
   
\caption{FIGURE 4. {\bf Left:} Schematic view of the regions having
similar time delay  for the radiation emitted by the source during a
short flare and scattered by the gas in the molecular clouds before
reaching the observer. Only clouds located along the surface of the
parabola will be bright in the reprocessed radiation at a given moment of time.
{\bf Right:} Morphology of the 6.4 keV line surface brightness
distribution as a 
function of time and relative position of the scattering cloud and the
compact source of continuum radiation. Time evolution assumes a short
flare of continuum emission. The time tags marked in 
the left column indicate the time elapsed since the moment,
when the surface of the illuminating parabola touched the cloud for the first
time. The
intensity of the the primary source was adjusted for each column in
order to have the same total flux from the cloud for the row marked 
`30'(years). The position of the
cloud with respect to the source is indicated by the pair of
numbers at the bottom of the figure. E.g. $(0;0)$ corresponds to the
source at the center of the cloud; $(-100;200)$ corresponds to the
cloud which is shifted by 100 pc to the left from the source and is
located 200 pc further away from the observer than the source.
Density distribution in the cloud adopted from Lis and Goldsmith, 1989.
} 
\label{geom}
\end{figure}

The simplest test (AXAF, XMM, ASTRO-E) would be to look for time
dependent changes of the surface brightness of the reprocessed
radiation (e.g. in the 
6.4 keV line, see Sunyaev \& Churazov, 1998). Indeed, for the short
flare the reprocessing sites 
(bright at a given moment of time after the primary flare) should be
located at the surface of the parabola with a focus at the position of
the primary source (Fig.4). All points over the surface of this
parabola have the same time delay for the photons coming from the
source to the reprocessing site and then to the observer. The parabola
will evolve with time effectively scanning the nonuniform distribution of the
scattering medium around the Sgr A*. The apparent motions could be sub
or superluminal, depending on the mutual location of the primary
source and the scattering site. Shown in Fig.4 are the time
dependent changes of the surface brightness of a spherical cloud at
different moments of time and different position of the cloud. For the
GC region one can expect ``proper motion'' of the regions bright in
the 6.4 keV line at a rate of the order of an arcminute per year (if
the rise and decay fronts of the flare were sharp enough).

\begin{figure}[t]
\hfil\psfig{file=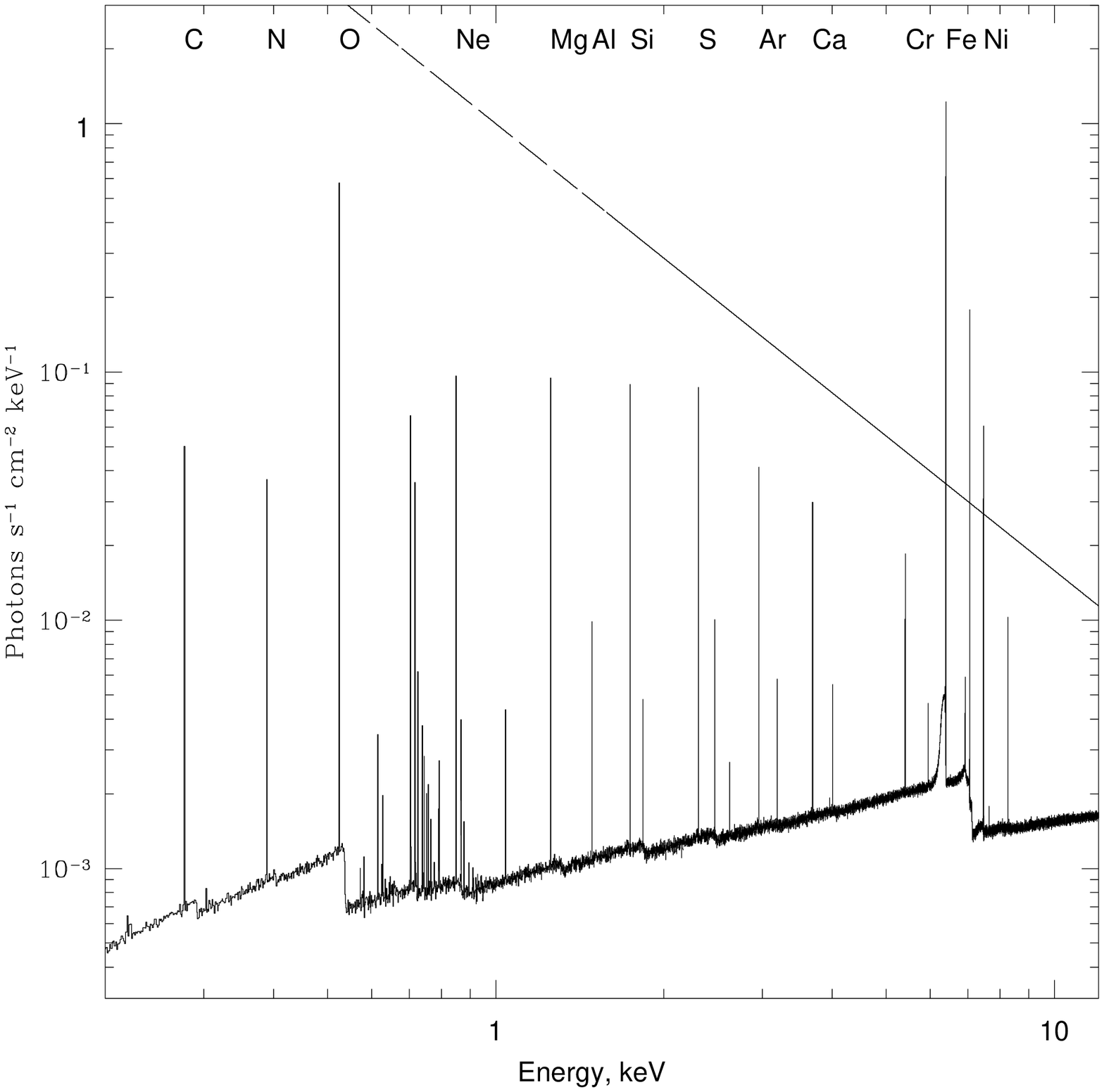,width=5.6cm}
\hfil\psfig{file=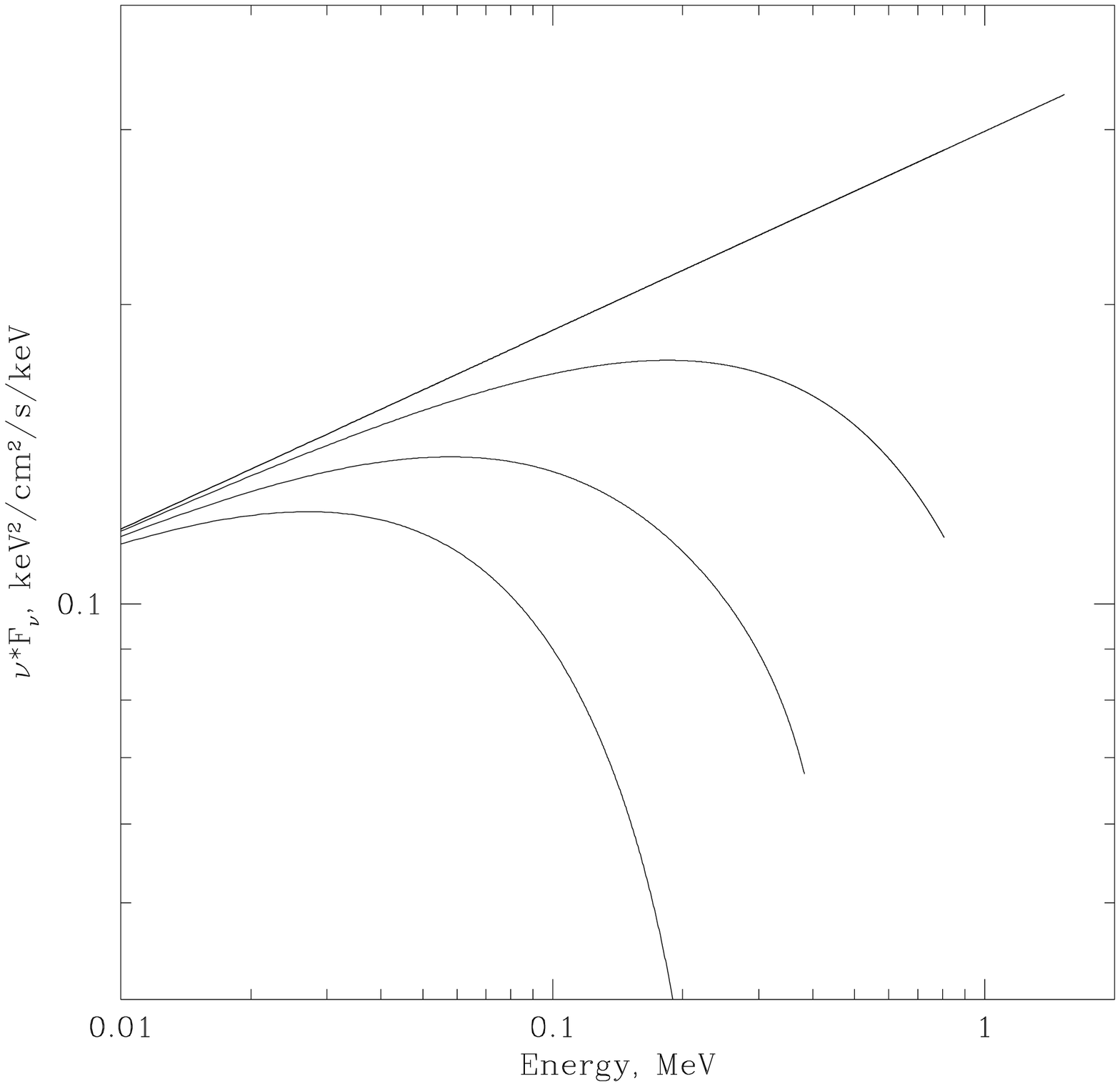,width=5.6cm}
   
\caption{FIGURE 5. {\bf Left:} Reflection spectrum for a semi-infinite
slab of neutral gas 
(hydrogen is in a molecular form) illuminated by a power law spectrum
(shown by dashed line). {\bf Right:} 
High energy part of the reprocessed spectra
(under assumption of an optically thin cloud) for the
different scattering angles (180\deg , 90\deg , 45\deg, 0\deg  from bottom to
top). The incident spectrum was assumed to be a power law with a photon
index of 1.8 extending up to an energy of 1.5 MeV. Normalization of the
reflected component depends on a solid angle occupied by the cloud and
its optical depth. 
} 
\label{refl}
\end{figure}

Important information can be obtained by studying the spectra of the
reflected component. Shown in Fig.5 is a spectrum reflected by a
semi-infinite slab of a molecular gas having solar abundance of heavy
elements. It is similar to the frequently used ``reflection'' model,
but fluorescent lines of the most abundant elements (not only iron
$K_\alpha$) are taken into account and electrons are assumed to be
bound in hydrogen molecules.  Measurement of precise energies of the 
fluorescent lines (ASTRO-E, Constellation, XEUS) will allow
direct comparison of the X--ray data and CO maps (velocity resolved)
in radio. It will be possible to unambiguously identify the clouds,
responsible for reprocessed radiation with the features in the CO maps. 
Note that electrons bound in molecules cause an increase of
the reflected continuum by a factor of 2 below 4-5 keV (due to
coherent scattering by electrons bound in hydrogen molecules) and
smearing of the ``Compton shoulder'' at the low energy side of the
bright emission lines (due to the ``motion'' of bound electrons within
molecules). 

One can try to identify the clouds already passed by a ``parabola''
of a primary radiation, using high sensitivity, high resolution
spectroscopy (Constellation, XEUS). Indeed, even although primary
photons already left the 
cloud, some fraction (of the order of Thomson optical depth of the
cloud) of the reprocessed photons will be ``reprocessed''
again and reach the observer after an additional delay of the order of
light crossing time of the cloud (we consider here the case of an
optically thin cloud). The flux seen by the observer will of course
decline with time, but the 
equivalent width of the fluorescent lines will on the contrary
increase (Sunyaev \& Churazov, 1998). The possibility of the detection of
such delayed emission may be severely limited by the very extended
low density envelopes of the clouds, leading to sufficiently strong
contribution of the reprocessed emission from the gas located further
away from observer and still exposed to the primary emission of the
source. 

If the spectrum of Sgr A* was hard and extended up to MeV, then
the shape of the continuum emission of the reflected component can be
used to determine the relative position of the scattering medium and
the primary source (Fig.5). Indeed in the limit of an optically thin cloud the
energy of a cut-off in the spectrum of the reflected component is
defined by the scattering angle, simply because the energy of high
energy photons after Compton scattering by a cold electron
strongly decreases if the scattering angle is large. Therefore a
detail study of the high energy continuum in the region with INTEGRAL
and correlation with the data in the standard X--ray band may help to
determine the relative position of the clouds with respect to the Sgr
A* and constrain the hard luminosity of the putative flare of Sgr A*.

\bsk
\baselineskip = 12pt
{\abstract \ni ACKNOWLEDGMENTS
This work was supported in part by RBRF grants 96-02-18588 and
97-02-16264. The SIGMA results are presented on behalf of the SIGMA
team (IKI, Moscow; SAp, Saclay; CESR, Toulouse).
}

\bsk
\baselineskip = 12pt


{\references \ni REFERENCES
\ssk
\ref Chen W., Gehrels N., Leventhal M., ApJ 1994, 426, 586
\ref Churazov E. et al., 1997, AdSpR, 19, 55
\ref Dubus G., Lasota J.-P., Hameury J.-M., Charles P.A., 1998, MNRAS,
in press (astro-ph/9809036)
\ref Eckart A., Genzel R., 1996, Nature, 383, 415
\ref Gilfanov M. et al., 1995, NATO  ASI Series, Series C, Vol.450
"The Lives  of  neutron  stars", ed.  A.Alpar, U.Kiziloglu, J.van
Paradijs, Kluwer Acad. Publishers, 331
\ref Goldwurm A. et al., 1994, Nature, 371, 589
\ref King A. et al., 1997a, ApJ, 484, 844
\ref King A., , Kolb U., Szuszkiewicz E., 1977b, ApJ, 488, 89
\ref Koyama K, 1994, New Horizon of X-ray Astronomy, FSS-12, 181, Univ. Acad. Press, Tokyo 
\ref Koyama K. et al., 1996, PASJ, 48, 249
\ref Kuznetsov S. et al., 1997, MNRAS, 292, 651
\ref Kuznetsov S. et al., 1998, Astronomy Letters, submitted
\ref Lis D.C. \& Goldsmith P.F., 1989, ApJ, 337, 704
\ref Lyuty V., Sunyaev R., 1976, Sov. Astron., 20, 290
\ref Marti' J. et al., 1998, A\&A Letters, in press
\ref Menou K., Narayan R. \& Lasota J.-P., 1998, ApJ, in press 
\ref Meyer F., Meyer-Hofmeister E., 1981, A\&A, 104., 10
\ref Predehl P, Truemper J., 1994, A\&A, 290, L29
\ref Shakura N., Sunyaev R., 1973, A\&A, 24, 337
\ref Sidoli et al., 1999 these proceedings
\ref Sunyaev R., Markevitch M., Pavlinsky M., 1993, ApJ, 407, 606 
\ref Sunyaev R., Churazov E., 1998, MNRAS, 297, 1279
\ref Uberini et al., 1999 these proceedings
\ref Van Paradijs J., 1996, ApJ Letters, 464, 139
}

\end{document}